\documentstyle[12pt]{article}

\begin{document}


\title{Chaotic Quantization of Classical Gauge Fields}

\author{
        T.S. Bir\'o$^1$, 
        S.G. Matinyan$^2$\thanks{U.~S.~address:
        3106 Hornbuckle Place, Durham, NC 27707}, 
        and 
        B. M\"uller$^3$
\\
\\
    $^1$MTA KFKI RMKI, H-1525 Budapest, P.O.Box 49, Hungary \\
    $^2$Yerevan Physics Institute, Yerevan, Armenia\\
    $^3$Department of Physics, Duke University, Durham, NC 27708-0305
}

\date{3 February 2001}

\maketitle

\begin{abstract}
We argue that the {\em quantized} non-Abelian gauge theory can be
obtained as the infrared limit of the corresponding {\em classical}
gauge theory in a higher dimension. We show how the transformation 
from classical to quantum field theory emerges, and calculate Planck's
constant from quantities defined in the underlying classical gauge
theory.
\end{abstract}



\section{Introduction}

Although much progress has been made in recent years, the question,
how gravitation and quantum mechanics should be combined into one
consistent unified theory of fundamental interactions, is still open. 
Superstring theory \cite{GSW}, which describes four-dimensional 
space-time as the low-energy limit of a ten- or eleven-dimensional 
theory (``M-theory'' \cite{Mth}), may provide the correct answer, 
but the precise form and content of the theory is not yet entirely 
clear. It is therefore legitimate to raise the question whether the 
fundamental description of nature at the Planck scale is really 
quantum mechanical, or whether the underlying theory could be a 
classical extension of general relativity. This questions was
initially raised by 't Hooft, who has argued that quantum mechanics
can logically arise as low-energy limit of a microscopically 
deterministic, but macroscopically dissipative theory \cite{tH99}.

It is the goal of this manuscript to present an explicit example
that shows how (Euclidean) quantum field theory can emerge in the
infrared limit of a higher-dimensional, classical field theory.
It is well known that relativistic quantum field theory in 
$(3+1)$-dimensional Minkowski space can be obtained by analytic 
continuation (``Wick rotation'') of the analogous statistical field 
theory defined on a four-dimensional Euclidean space. In fact, this 
concept provides the only known mathematically rigorous definition 
of interacting quantum field theories. Physical observables, such 
as vacuum expectation value of self-adjoint operators, can be 
reliably calculated in the Euclidean path integral formulation
of the quantum field theory. This method has been extensively used 
to obtain numerical solutions of many relativistic quantum field 
theories.

We here show that in some cases, specifically for non-Abelian gauge
fields, the functional integral of the three-dimensional Euclidean
{\em quantum} field theory arises naturally as the long-distance limit
of the corresponding {\em classical} gauge theory defined in
$(3+1)$-dimensional Minkowski space. Because of the general nature of 
the mechanism underlying this transformation, for which we have coined 
the term {\em chaotic quantization}, it is expected to work equally well 
in other dimensions. For example, the four-dimensional Euclidean quantum 
gauge theory arises as the infrared limit of the $(4+1)$-dimensional 
classical gauge theory. We emphasize that the dimensional reduction is
not caused by compactification; the classical field theory does not
exhibit periodicity either in real or imaginary time.

The mechanism discussed below can be viewed as a physical analogue of 
the well-known technique of stochastic quantization \cite{PW81}. In both
cases, the quantum fluctuations arise from the stochastic noise in
a higher-dimensional theory. However, chaotic quantization differs 
from stochastic quantization in two essential aspects: It only applies
to certain field theories, including non-Abelian gauge theories, and
it allows us to {\em calculate} Planck's constant $\hbar$ in terms of
fundamental physical quantities of the underlying higher-dimensional
classical field theory. Accordingly, chaotic quantization provides
a {\em physical} mechanism generating the quantum mechanics of fields 
and particles, while stochastic quantization is generally regarded as 
a convenient calculational technique, but not as a physical principle.

\section{Chaoticity of Classical Yang-Mills Fields}

The chaotic nature of classical non-Abelian gauge theories was first
recognized twenty years ago \cite{MST81,BMM95}. Over the past decade, 
extensive numerical solutions of spatially varying classical non-Abelian 
gauge fields on the lattice have revealed that the gauge field has 
positive Lyapunov exponents that grow linearly with the energy density 
of the field configuration and remain well-defined in the limit of small 
lattice spacing $a$ or weak-coupling \cite{MT92,Go93,Mu96}. More recently, 
numerical studies have shown that the $(3+1)$-dimensional classical 
non-Abelian lattice gauge theory exhibits global hyperbolicity.
This conclusion is based on calculations of the complete spectrum 
of Lyapunov exponents \cite{Go94} and on the long-time statistical 
properties of the Kolmogorov-Sinai (KS) entropy of the classical SU(2) 
gauge theory \cite{BMS99}. 

These results imply that correlation functions of physical observables 
decay rapidly, and that long-time averages of observables for a single 
initial gauge field configuration are identical to their microcanonical 
phase-space average, up to Gaussian fluctuations which vanish in the 
long-time limit as $t_s^{-1/2}$, where $t_s$ is the observation time. 
Since the relative fluctuations of extensive quantities scale as 
$L^{-3/2}$, the microcanonical (fixed-energy) average can be safely 
replaced by the canonical average when the spatial volume probed by
the observable becomes large. In the following we discuss the hierarchy
of time and length scales on which this transformation occurs.

Accoridng to the cited results, the classical non-Abelian gauge field 
self-thermalizes on a finite time scale $\tau_{\rm eq}$ given by the 
ratio of the equilibrium entropy and the KS-entropy, which determines
the growth rate of the course-grained entropy: 
\begin{equation}
\tau_{\rm eq} = S_{\rm eq}/h_{\rm KS}\, .
\end{equation}
At weak coupling, the KS-entropy for the $(3+1)$-dimensional
SU(2) gauge theory scales as
\begin{equation}
h_{\rm KS} \sim g^2 E \sim g^2 T (L/a)^3\, ,
\end{equation}
where $E$ is the total energy of the field configuration and $T$ is
the related temperature defined by 
\begin{equation}
E = T^2 \partial Z/\partial T\, . 
\end{equation}
Here $Z(T)$ is the partition function of the classical gauge field
regularized by the lattice cut-off $a$. The equilibrium entropy of the 
lattice is independent of the energy and proportional to the number 
of degrees of freedom of the lattice: $S_{\rm eq} \sim (L/a)^3$. The time
scale for self-equilibration is thus given by\footnote{In the convention 
adopted here, $g$ is the coupling strength of the classical 
Yang-Mills theory with dimension (energy$\times$length)$^{-1/2}$. 
This choice ensures the proper dimensionality of the Yang-Mills action.} 
\begin{equation}
\tau_{\rm eq} \sim (g^2 E a^3/L^3)^{-1} \sim (g^2 T)^{-1}\, .
\end{equation}

When one is interested only in long-term averages of observables,
it is thus sufficient to consider the {\em thermal} classical gauge
theory on a three-dimensional spatial lattice. Note that, although
we are interested only in the long-distance properties of the 
quasi-thermalized field, we have to define the classical gauge field 
on a lattice rather than in the continuum. Due to the nonlinear 
interactions most of the energy contained in the initial field 
configuration ultimately cascades into modes with wave lengths 
near the ultraviolet cutoff $a$, and the limit of vanishing lattice 
spacing $a$ is not well defined. Without the lattice cutoff, we would 
be unable to replace the microcanonical average by a thermal average,
because the cascade toward the ultraviolet would not end and no
stationary limit would exist. We shall see later that the lattice 
cutoff $a$ also takes on an important physical role, as it enters 
into the definition of Planck's constant $\hbar$.

\section{Chaoticity of Classical Yang-Mills Theory}

The long-distance dynamics of non-Abelian gauge theories at finite 
temperature $T$ is known to reduce to the dynamics of the static 
chromomagnetic sector of the gauge field \cite{Na83}. The scale 
beyond which this dimensional reduction is valid is given by the 
{\em magnetic length scale} $d_{\rm mag} \sim (g^2T)^{-1}$, where 
$g$ is the classical gauge coupling.
The magnetic length scale is the same for the classical 
and the quantized gauge theories at finite temperature. 
$d_{\rm mag}$ is independent of the lattice cutoff $a$. It is well 
recognized that the static magnetic sector in the thermal quantum 
field theory is essentially classical in nature and depends on 
$\hbar$ only via the scale of the thermal effective gauge 
coupling $g(T)$.

It is worth noting that in spite of many similarities between the
thermal classical field theory and the thermal quantum field theory,
there are major differences. The ultraviolet properties of the 
quantum field theory at finite temperature are controlled by the 
thermal length scale $d_{\rm th} \sim \hbar/T$, which is a 
basically quantum mechanical concept.\footnote{The need for this 
length scale in the derivation of the Stefan-Boltzmann radiation law 
motivated Planck in 1900 to postulate the existence of the quantum 
of action $\hbar$.} In the thermal classical field theory, the lattice 
spacing $a$ serves as ultraviolet regulator. Similarly, the electric 
screening length of the thermal quantum field theory is
$d_{\rm el} \sim \sqrt{\hbar}/gT$. In the thermal classical field 
theory electric fields are screened on the length scale 
$d_{\rm el} \sim \sqrt{a/g^2T}$. Only the magnetic length scales are 
equal for classical and quantal gauge fields. The inverse electric 
screening length is proportional to the plasma frequency 
$\omega_{\rm pl}$ governing propagating long-wavelength modes. 
The damping of these plasma modes is of the order
$\gamma_{\rm pl} \sim d_{\rm mag}^{-1}$ \cite{BP90}, 
rendering the dynamics of the thermal gauge field purely 
dissipative and noisy on distances larger than $d_{\rm mag}$.

The dynamic properties of thermal non-Abelian gauge fields at such
long distances have been studied in much detail \cite{Bo98,ASY99,ASY97}. 
It is now understood that the real-time dynamics of the gauge field at 
such scales can be described, at leading order, by a Langevin equation
\begin{equation}
\sigma {\partial A\over\partial t} = - D \times B + \xi\, ,
\label{LANGEVIN}
\end{equation}
where $D$ is the gauge covariant spatial derivative, $B = D \times A$ 
is the magnetic field strength, and $\xi$ denotes Gaussian distributed 
(white) noise with zero mean and variance
\begin{equation}
\langle \xi_i(x,t)\xi_j(x',t') \rangle
= 2\sigma T \delta_{ij}\delta^3(x-x')\delta(t-t')\, .
\label{NOISE}
\end{equation}
Here $\sigma$ denotes the color conductivity \cite{GS93} of the thermal 
gauge field which is determined by the ratio 
$\omega_{\rm pl}^2/\gamma_{\rm th}$  of the plasma frequency 
$\omega_{\rm pl}$ and the damping rate $\gamma_{\rm th}$ 
of a thermal gauge field excitation.

At leading logarithmic order in the quantum field theory, the color
conductivity satisfies
\begin{equation}
\sigma^{-1} \sim \frac{\hbar}{T} \ln[d_{\rm mag}/d_{\rm el}] \, .
\label{SIGQ}
\end{equation}
The derivation of the Langevin equation (\ref{LANGEVIN}) for
the classical thermal gauge theory proceeds completely in
parallel to that for the quantum field theory, except that the
plasma frequency $\omega_{\rm pl}^2 \sim g^2T/a$, and the ratio 
between the magnetic and electric length scales depends 
on the combination $g(Ta)^{1/2}$ instead of $g\hbar^{1/2}$.
The separation of length scales requires $g^2Ta \ll 1$.
The color conductivity then scales as
\begin{equation}
\sigma^{-1} \sim a \ln[d_{\rm mag}/d_{\rm el}] \, .
\label{SIGC}
\end{equation}
This relation implies that the color conductivity is an ultraviolet
sensitive quantity, which depends on the lattice cutoff. The various
relations derived in this section are summarized in Table 
\ref{tab:class-qm}, where the results for the thermal quantum field
theory are listed in parallel with those of the classical gauge 
theory.

\section{Dimensional Reduction and Quantization}

We will now show that an observer restricted to long distances in 
three-dimensional Euclidean space would interpret the dynamics of 
the classical gauge field as that of a quantum field in its vacuum 
state with the Planck constant 
\begin{equation}
\hbar_3 = aT \, .
\label{hbar}
\end{equation}
The starting point of our argument is the well-established fact that the
random Gaussian process defined by the Langevin equation (\ref{LANGEVIN}) 
generates three-dimensional field configurations with a probability 
distribution $P[A]$ determined by the Fokker-Planck equation 
\begin{equation}
\sigma\,\frac{\partial}{\partial t} P[A] \, = \,
\int \! d^3x \, \frac{\delta}{\delta A} \left( T \frac{\delta P}
{\delta A} + \frac{\delta W}{\delta A}\, P[A] \right) \, .
\label{FOKKER}
\end{equation}
Here $W[A]$ denotes the magnetic energy functional
\begin{equation}
W[A] = \int\! d^3x\, \frac{1}{2} B(x)^2 \, .
\label{WA}
\end{equation}
Any non-static excitations of the magnetic sector of the gauge field,
i.e. magnetic fields $B(k)$ not satisfying $k\times B=0$, die away 
rapidly on a time scale of order $\sigma/k^2$, where $k$ denotes the 
wave vector of the field excitation. For observers sensitive only 
to distances much larger than $d_{\rm mag}$ and times much longer
than $\sigma d_{\rm mag}^2$, measurements of the magnetic field 
yield averages with the equilibrium weight given by the stationary 
solution of the Fokker-Planck equation (\ref{FOKKER}):
\begin{equation}
P_0[A] = e^{-W[A]/T}\, .
\label{PA}
\end{equation}
The three-vector $B_i \sim \epsilon_{ijk}F^{jk}$ incorporates all 
components of the field strength tensor $f^{jk}$ in three dimensions. 
$W/T$ is now identified as the three-dimensional action 
$S_3$ measured in units of Planck's constant $\hbar_3$:
\begin{equation}
W/T = S_3/\hbar_3\, ,
\label{LOWDIM}
\end{equation}
where
\begin{equation}
S_3[A] = - \frac{1}{4} \int\! dx_3 \int\! d^2x\, f^{ik}f_{ik}\, .
\label{SA}
\end{equation}
Dimensional reasons require a rescaling of the gauge field strength 
with the fundamental length scale
\begin{equation}
f^{ik} = \sqrt{a} F^{ik}\, .
\label{FSCALE}
\end{equation}
That the lattice spacing $a$ is the proper rescaling parameter is 
seen by noting that the lattice versions of the two-dimensional 
and the three-dimensional integrals 
\begin{equation}
\int d^nx \quad\to\quad a^n\sum_x
\end{equation}
differ by a factor $a$. Together with the relations (\ref{WA}), 
(\ref{LOWDIM}), and (\ref{SA}) this reasoning demonstrates that 
$\hbar_3 = aT$, as stated at the beginning of this section. The 
rescaling of the gauge field also fixes the three-dimensional 
coupling constant
\begin{equation}
g_3^2 = \frac{g^2}{a} = \frac{g^2T}{\hbar_3} \, ,
\label{G3}
\end{equation}
so that
\begin{equation}
\sqrt{a}(\partial A + g A\times A) =
(\partial A_{(3)} + g_3 A_{(3)}\times A_{(3)})\, .
\end{equation}

According to (\ref{FOKKER}), an observer confined to the measurement
of long-time and long-distance averages of microscopic observables
associated with the classical gauge field measures the same values 
as would an observer ``living'' in the three-dimensional Euclidean
world in the presence of a quantized gauge field in its vacuum state.
It is important to note that this correspondence is not induced by
a compactification of the Minkowskian time coordinate. There is no 
true thermal bath of gauge fields in the original Minkowski space 
theory, and the quasi-thermal solution of the $(3+1)$-dimensional 
classical field theory does not satisfy periodic boundary conditions 
in imaginary time. 

The effective dimensional reduction found here is not caused by 
the discreteness of the excitations with respect to the time-like 
dimension, but by the dissipative nature of the (3+1)-dimensional 
dynamics. Magnetic field configurations satisfying $D\times B=0$ 
can be thought of as low-dimensional attractors of the dissipative 
motion, and the chaotic dynamical fluctuations of the gauge field 
around the attractor can be consistently interpreted as quantum 
fluctuations of a vacuum gauge field in 3-dimensional Euclidean
space. 

We thus see that the mechanism of dimensional reduction discussed 
above is distinct from the mechanism that operates in thermal
quantum field theories. In fact, the dimensional reduction by 
chaotic fluctuations and dissipation does not occur in scalar
field theories, because -- even in cases that exhibit chaos, such 
as two quarticly coupled scalar fields -- there is no dynamical 
sector that survives after long-time averaging. Quasi-thermal 
fluctuations generate a dynamical ``mass'' for the scalar field(s) 
and thus eliminate any arbitraily slow field modes.\footnote{An
exception may be the case where the excitation energy of the scalar
field is just right to put the quasi-thermal field at the critical
temperature of a second-order phase transition, where arbitraily
slow modes exist as fluctuations of the order parameter.} In the 
case of gauge fields, the transverse magnetic sector is protected 
by the gauge symmetry, and it is this sector which survives the
time average, without any need for fine-tuning of the microscopic
theory.

\section{General Considerations}

It is worthwhile to review the essential ingredients of chaotic 
quantization. First, the underlying classical field theory must 
contain strongly coupled massless degrees of freedom. Such theories 
are generally chaotic at the classical level \cite{MM97}.  When 
observations are restricted to the infrared degrees of freedom, 
this corresponds to a coarse graining of the dynamical system, 
leading to strongly dissipative long-distance dynamics. The coupling
to the short-distance modes generates uncorrelated noise, and the 
coarse-grained system obeys a dissipation-fluctuation theorem
\cite{GM96}.  

Second, the classical field theory at finite temperature must have 
degrees of freedom that remain unscreened. This condition generally 
requires the presence of a symmetry, such as gauge invariance. 
It is a reasonable expectation that such symmetries occur in any 
unified theory containing general relativity. The requirement also
provides a simple and consistent explanation for the empirical fact
that all fundamental interactions are described by gauge fields. 

Our example for the chaotic quantization of a three-dimensional 
gauge theory in Euclidean space raises a number of questions:
\begin{enumerate}
\item
Does the principle of chaotic quantization generalize to higher
dimensions, in particular, to quantization in four dimensions?
\item
Can the method be extended to describe field quantization in 
Minkowski space?
\item
What type of deviations from the standard quantum field theory 
are caused by the existence of a microscopic classical dynamics?
\end{enumerate}

The first question is most easily answered. As long as globally
hyperbolic classical field theories can be identified in higher
dimensions, our proposed mechanism should apply. Although we do
not know of any systematic study of dicretized field theories in
higher dimensions, a plausibility argument can be made that
Yang-Mills fields exhibit chaos in (4+1) dimensions. For this
purpose, we consider the infrared limit of a spatially constant
gauge potential, as studied in refs. \cite{MST81,BMM95,AS83}. For 
the SU($N$) gauge field in $(D+1)$ dimensions in the $A_0=0$ gauge, 
there are $3(N^2-1)$ interacting components of the vector potential 
and $3(N^2-1)$ canonically conjugate momenta (the components of the 
electric field) that depend only on the time coordinate. The 
remaining gauge transformations and Gauss' law allow to eliminate 
$(N^2-1)$ degrees of freedom from each. Next, rotational invariance 
in $D$ dimensions permits to reduce the number of dynamical degrees 
of freedom by twice the number of generators of the group SO($D$), 
i.e. by $D(D-1)$. This leaves a $(D-1)(2N^2-2-D)$-dimensional phase 
space of the dynamical degrees of freedom and their conjugate 
momenta. For the dynamics to be chaotic, this number must be at 
least three. For the simplest gauge group SU(2), this condition
permits infrared chaos in $2\le D\le 5$ dimensions, including the
interesting case $D=4$. Higher gauge groups are needed to extend 
the chaotic quantization scheme to gauge fields in $D>5$ dimensions.
Of course, this reasoning does not prove full chaoticity of the
Yang-Mills field in these higher dimensions, it just indicates the
possibility. Numerical studies will be required to establish the
presence of strong chaos in these classical field theories.

The second question is more difficult to address. A formal answer would
be that the Minkowski-space quantum field theory can (and even must)
be obtained by analytic continuation from the Euclidean field theory.
Any observable in the Minkowski-space theory that can be expressed 
as a vacuum expectation value of field operators can be obtained in 
this manner. If this argument appears somewhat unphysical,
one might consider a completely different approach, beginning with
a chaotic classical field theory defined in (3+2) dimensions. 
Field theories defined in spaces with
two time-like dimensions were first proposed by Dirac in the context
of conformal field theory \cite{Dirac} and have recently been 
considered as generalizations of superstring theory \cite{Bars}.
In that case, the reduction to one time dimension is achieved by
gauge fixing. In the present case, the physical time dimension may
be defined as the coordinate orthogonal to the total 5-momentum
vector $P^\mu$ of the initial field configuration.\footnote{In the
four-dimensional case, the total 4-momentum vector $P^\mu$, which
is assumed to be time-like, defines the 4-velocity vector $u^\mu$ 
of the thermal rest frame via the relation $P^\mu = E(T) u^\mu$.
The three-dimensional Euclidean quantum field theory lives in the
hypersurface orthogonal to $u^\mu$.}

In the presence of two time-like dimensions, ``energy'' becomes a
two-component vector $\vec E$ that is a part of the $(D+2)$-dimensional
energy-momentum vector. If we select an initial field configuration 
with energy $E_0{\vec n}$, where $\vec n$ is a two-dimensional unit
vector, this choice defines a preferred time-like direction $\vec n$ 
in which the field thermalizes. Conservation of the energy-momentum
vector ensures that the total energy component orthogonal to $\vec n$
always remains zero. In this sense, the choice of an initial field
configuration corresponds to a spontaneous breaking of the global
SO($D$,2) symmetry down to a global SO($D$,1) symmetry. Whether this
leads to an effective quantum field theory in $(D+1)$ dimensional
Minkowski space, remains to be investigated.

Finally, it is interesting to ask which deviations from the quantum
field theory could be detected by a ``slow'' observer by means of
very precise measurements. Clearly, an observer able to resolve 
the dynamics on the thermal or electric length scales of the 
underlying classical field theory would observe deviations from
the dimensionally reduced vacuum field theory. For a space-time
volume of linear dimension $L$, the amplitude of the fluctuations
is of the order $(g^2TL)^{-2}$. If, as one might suspect, the 
relevant microscopic length scale is of the order of the Planck
scale, $g^2T \sim M_P$, quantities sensitive to the fluctuations
around the infrared dynamics are suppressed by $(M_P L)^2$. For
presently accessible length scales, the suppression factor is
at least $10^{-34}$, and even smaller in low-energy precision
tests of quantum mechanics. However, in principle, tests of 
Bell's inequality in systems prepared with strong correlations 
on short time and distance scales can be used to establish at 
least an upper bound for the scale at which the transition from
the classical dissipative dynamics to the quantum dynamics occurs.

It is a natural question to ask whether the mechanism of chaotic 
quantization outlined above corresponds to a hidden parameter
theory of quantum mechanics. The answer is obviously positive,
as the microscopic state of the system in a higher dimension
is always precisely and deterministically defined. However, it
is important to realize that the impossibility of hidden parameter 
descriptions of quantum mechanics is restricted to {\em local} 
theories, while our proposed mechanism operates in a higher
dimensional space. A local dynamical theory in more dimensions
generates fundamentally non-local effects in the lower-dimensional 
space. One can speculate that the time-scale associated with 
dimensional reduction, $(g^2T)^{-1}$, is the time for the collapse 
of the wave function. Our analysis predicts that this is the time 
required to average over the noise generated by the classical
dynamics and to establish the stationary distribution of the 
Fokker-Planck equation (\ref{FOKKER}).

\section{Summary and Conclusions}

Let us summarize. We have shown that a homogeneously excited, 
classical field theory in four dimensions can generate a 
three-dimensional Euclidean quantum field theory. Whereas the
classical theory appears thermal for a four-dimensional
observer, the three-dimensional observer experiences quantum 
fields at zero temperature. The essential feature facilitating 
this transformation is that the underlying deterministic theory 
contains a mechanism for information loss \cite{tH99,tH88,tH97}, 
here realized through its chaotic dynamics.

We can go further and speculate that the randomness caused
by this intrinsic chaoticity of the underlying theory could
generally lead to a reduction of the effective space-time
dimensionality of the theory. In our example, the dimensional
reduction is an effect of the quasi-thermal fluctuations.
Another related, well-known phenomenon is the dimensional
reduction of spin systems in arbitrarily weak, random magnetic 
fields \cite{IM75}, which finds its explanation in the hidden
supersymmetry of the system \cite{PS79}.

We note that symmetries and physical laws may arise naturally 
from some essentially random dynamics, rather than being 
postulated from the beginning \cite{NN78,FNN80,FN91}. The goal 
of the program formulated in our example, is more restricted:
microscopic randomness is utilized as a foundation for 
``large-scale'' physics that is described by quantum mechanics.

It is not clear whether the mechanism presented here for
non-Abelian gauge fields is also at work in general relativity.
Examples of chaotic behavior have been identified in the 
dynamics of classical gravitational fields \cite{NATO,Grav}.
It has been found that the chaotic nature of the solutions may
depend on the number of dimensions. A famous case is the 
evolution toward the singularity in the Bianchi type IX
geometry, where the chaotic oscillatory approach changes to a
monotonic approach in more than 10 dimensions \cite{DRH89,Ma00}.

It has not been demonstrated that chaoticity is a general property
of solutions of Einstein's equations. This may not even be required,
because an entirely different mechanism of information loss may be 
at work in general relativity than in Yang-Mills theory. Indeed, 
't Hooft has speculated that black hole formation may be the 
mechanism operating in the case of gravity \cite{tH99}. 
Finally, our example does not contain fermion fields. 
It would be interesting to extend our study to supersymmetric 
theories, some of which have been shown to exhibit chaos in 
the infrared limit \cite{AKM98}.

\bigskip
\noindent {\em Acknowledgments:}
This work was supported in part by a grant from the
U.S. Department of Energy (DE-FG02-96ER40495), 
by the American-Hungarian Joint Fund T\'ET (JFNo. 649)
and by the Hungarian National Science Fund OTKA (T 019700).
One of us (B.M.) acknowledges the support by a 
U.~S.~Senior Scientist Award from the Alexander von Humboldt
Foundation.

\section*{Appendix}

Here we present a qualitative analysis of the various length and
time scales of a thermal Yang-Mills field in $D=(d+1)$ space-time
dimensions, for $d\geq 2$. The results are accurate up to
logarithmic corrections that arise for various quantities in
some dimensions. We decompose the field into Fourier components 
(suppressing vector and color indices):
\begin{equation}
A(x,t) = V^{1/2} \int d^dk a(k) e^{ikx-i\omega_k t} \, .
\label{A1}
\end{equation}
Equipartition of the energy over the classical modes then implies
that
\begin{equation}
\vert a(k)\vert^2 = T \omega_k^{-2} \, .
\label{A2}
\end{equation}
The presence of an ``external'' static color potential $A^0$
induces a polarization density
\begin{equation}
\rho_{\rm pol} \sim g^2 V^{-1} \int d^dx A(x,t)^2 A^0
\sim g^2 A^0 \int d^dk \vert a(k)\vert^2 
\sim g^2T A^0 \int d^dk \omega_k^{-2} \, .
\label{A3}
\end{equation}
With the ultraviolet lattice cut-off $k\leq a^{-1}$, one obtains:
\begin{equation}
d_{\rm el}^{-2} = \rho_{\rm pol}/A_0 \sim g^2Ta^{2-d} \, .
\end{equation}
The field theory can be considered to in the weak coupling
regime, when the electric screening length is much larger than
the lattice constant $a$. This amounts to the condition
\begin{equation}
{\tilde g}^2 \equiv g^2Ta^{4-d} = (a/d_{\rm el})^2 \ll 1 \, ,
\label{A4}
\end{equation}
which defines the effective weak coupling parameter ${\tilde g}$
of the $(d+1)$-dimensional classical Yang-Mills theory. With the
help of this parameter, the electric screening length can be
expressed simply as $d_{\rm el} = a/{\tilde g}$.

The coupling constant of the dimensionally reduced quantum field
theory in $D-1=d$ dimensional Euclidean space is given by
\begin{equation}
g_{D-1}^2\hbar = g^2a^{-1}(Ta) = g^2T \, ,
\label{A5}
\end{equation}
independent of the number $d$ of space dimensions. 

The transport coefficient describing color conductivity
$\sigma$ is obtained by considering the polarization current
induced by a constant electric field $E$. Schematically, it 
is given by
\begin{equation}
j_{\rm pol} = g^2 V^{-1} \int d^dxdt A(x,t)^2 E
\sim g^2 E \int d^dk \vert a(k)\vert^2 \gamma(k)^{-1} \, ,
\label{A6}
\end{equation}
where $\gamma(k)$ is the damping rate of a thermal field mode.
This damping rate can be calculated in the standard way using
the formula $\gamma = \sigma_{\rm coll} n_{\rm th}$, where
$\sigma_{\rm coll}$ denotes the ``cross section'' for a thermal
excitation, and $n_{\rm th}$ stands for the density of hard
thermal excitations. In $d$ spatial dimensions one finds
\begin{equation}
\sigma_{\rm coll} 
\sim g^4Ta \int d^{d-1}q (q^2+d_{\rm el}^{-2})^{-2}
\sim g^4Ta d_{\rm el}^{5-d} \, .
\label{A7}
\end{equation}
The classical formula contains an additional factor $(Ta)$
describing the enhancement due to the classical occupation
of thermal modes of the gauge field. The density of thermal
excitations is
\begin{equation}
n_{\rm th} \sim \int d^dk \vert a(k)\vert^2 \omega_k
\sim T a^{1-d} \, .
\label{A8}
\end{equation}
Since $\gamma$ does not depend on $k$ in this approximation,
we can pull it out of the integral over $k$ in (\ref{A6})
to obtain:
\begin{equation}
j_{\rm pol} \sim g^2 E\gamma^{-1} \int d^dk \vert a(k)\vert^2 
\sim g^2 E\gamma^{-1} T a^{2-d} \, .
\label{A9}
\end{equation}
Combining the expressions (\ref{A7}--\ref{A9}) we final get
the desired expression for the color conductivity:
\begin{equation}
\sigma = j_{\rm pol}/E \sim g^2a/\sigma_{\rm coll} 
\sim d_{\rm el}^{d-5} (g^2T)^{-1} \, .
\label{A10}
\end{equation}
Specifically, in $d=4$ spatial dimensions, the result is
\begin{equation}
\sigma \sim a (g^2T)^{-3/2} \, .
\end{equation}

All results are summarized, and compared to the results
obtained in the $(d+1)$-dimensional thermal quantum field
theory, in Table \ref{tab:d-dim}.


\begin{table}[ht]
\caption{Comparison of length scales in the classical and
 quantum Yang-Mills theories: hard thermal scale $d_{\rm th}$,
 electric scale $d_{\rm el}$, magnetic scale $d_{\rm mag}$,
 plasma frequency $\omega_{\rm pl}$, damping rate $\gamma_{\rm th}$, 
 and color conductivity $\sigma$.}
\label{tab:class-qm}
\vspace{0.2cm}
\begin{center}
\begin{tabular}{|l|c|c|}
\hline
{} & Quantum field theory & Classical field theory
\\
\hline
 $d_{\rm th}$  & $\hbar T^{-1}$         & $a$               \\
\hline
 $d_{\rm el}$  & $\hbar^{1/2}(gT)^{-1}$ & $(g^2T/a)^{-1/2}$ \\
\hline
 $d_{\rm mag}$ & $(g^2T)^{-1}$          & $(g^2T)^{-1}$     \\
\hline
 $d_{\rm mag} \gg d_{\rm el}$  
               &  $g^2\hbar \ll 1$      & $g^2Ta \ll 1$     \\
\hline
 $\omega_{\rm pl}^2$  
               & $(gT)^2/\hbar$         & $g^2T/a$          \\
\hline
 $\gamma_{\rm th}$    
               & $g^2T\ln[d_{\rm mag}/d_{\rm el}]$ 
                        & $g^2T\ln[d_{\rm mag}/d_{\rm el}]$ \\
\hline
 $\sigma$      & $\omega_{\rm pl}^2/\gamma_{\rm th}$ 
                      & $\omega_{\rm pl}^2/\gamma_{\rm th}$ \\
\hline
\end{tabular}
\label{class-qm}
\end{center}
\end{table}

\hskip2cm

\begin{table}[ht]
\caption{Characteristic scales for classical and quantized 
 thermal Yang-Mills theories in $D=(d+1)$ space-time dimensions:
 weak coupling parameter ${\tilde g}$, electric screening length
 $d_{\rm el}$, magnetic length scale $d_{\rm mag}$, coupling
 constant of the dimensionally reduced quantum field theory
 $g_{D-1}$, and color conductivity $\sigma$.}
\label{tab:d-dim}
\vspace{0.2cm}
\begin{center}
\begin{tabular}{|l|c|c|}
\hline
 {} & Quantum field theory & Classical field theory \\
\hline
 ${\tilde g}^2$ & $g^2T^{d-3}\hbar^{4-d}$ & $g^2Ta^{4-d}$    \\
\hline
 $d_{\rm el}$   & $\hbar/({\tilde g}T)$   & $a/{\tilde g}$   \\
\hline
 $d_{\rm mag}$  & $\hbar/({\tilde g}^2T)$ & $a/{\tilde g}^2$ \\
\hline
 $g_{D-1}^2$    & $g^2T/\hbar$            & $g^2/a$          \\
\hline
 $\sigma$       & $d_{\rm el}^{d-5} (g^2T)^{-1}$ 
                          & $d_{\rm el}^{d-5} (g^2T)^{-1}$ \\
\hline
\end{tabular}
\label{d-dim}
\end{center}
\end{table}


\begin{thebibliography}{99}
\bibitem{Mth}
M-theory reference, to be filled in.
\end{thebibliography}
\end{document}